\documentclass[12pt,a4paper]{article}
\usepackage{latexsym,amsmath,graphicx,epsfig,amssymb,rotating}
\usepackage{amssymb}

\newcommand{\mt}{t\kern-0.035cm\char39\kern-0.03cm}
\newcommand{\ml}{l\kern-0.035cm\char39\kern-0.03cm}
\newcommand{\md}{d\kern-0.035cm\char39\kern-0.03cm}
\newcommand{\mL}{L\kern-0.035cm\char39\kern-0.03cm}
\newcommand\chapter{\pagestyle{myheadings}}
\newtheorem{ex1}{Example }
\newtheorem{lem}{Lemma}
\newtheorem{prop}{Proposition}
\newtheorem{coro}{Corollary}

\newtheorem{rem}{Remark}

\begin{document}

\title{Bell-type Inequalities for Bivariate Maps on Orthomodular Lattices}
\author{Jaros{\l }aw Pykacz\thanks{Institute of Mathematics, University of Gda\'nsk, Poland; e-mail: pykacz@mat.ug.edu.pl},  \mL ubica Val\'{a}\v{s}kov\'{a}\thanks{Department of Mathematics and Descriptive Geometry, Faculty of Civil Engineering, Slovak University of Technology in Bratislava, Slovakia; e-mail: lubica.valaskova@stuba.sk}, O\ml ga N\'{a}n\'{a}siov\'{a}\thanks{Institute of Computer Science and Mathematics,
Faculty of Electrical Engineering and Information Technology, Slovak
University of Technology in Bratislava, Slovakia; e-mail:
olga.nanasiova@stuba.sk}}
\date{}
\maketitle
\thispagestyle{empty}

\begin{abstract}

Bell-type inequalities on orthomodular lattices, in which conjunctions of propositions are not modeled by meets but by maps for simultaneous measurements ($s$-maps), are studied. It is shown, that the most simple of these inequalities, that involves only two propositions, is always satisfied, contrary to what happens in the case of traditional version of this inequality in which conjunctions of propositions are modeled by meets. Equivalence of various Bell-type inequalities formulated with the aid of bivariate maps on orthomodular lattices is studied. Our invesigations shed new light on the interpretation of various  multivariate maps defined on orthomodular lattices already studied in the literature. The paper is concluded by showing the possibility of using $s$-maps and $j$-maps to represent counterfactual conjunctions and disjunctions of non-compatible propositions about quantum systems.


\end{abstract}
\section{INTRODUCTION}

Lindenbaum-Tarski algebras of theories in which propositions obey the rules of classical logic are Boolean algebras (BAs). In particular, meet $(\wedge $) in these Boolean algebras describes conjunction of propositions. The same is usually assumed in the case of orthomodular lattices (OMLs) which are non-distributive generalizations of BAs, also in the case of lattices $\cal L(\cal H)$ of projections onto closed linear subspaces of Hilbert spaces $\cal H$ used to describe quantum systems. Only rarely it is noticed that careless interpretation of meets in OMLs as always representing conjunction of propositions may lead to difficulties (see, e.g., Jammer 1974). Impossibility of simultaneous verification (non-compatibility) of some propositions about a physical system is a remarkable feature of quantum systems. However, in OMLs meet is a global operation, therefore it can be applied also to non-compatible elements of these lattices, which in the case of Hilbertian lattices  $\cal L(\cal H)$ makes its interpretation as a conjunction of propositions doubtful from the physical point of view.

Bell-type inequalities, studied intensively also in the realm of OMLs (Santos 1986; Pykacz 1989; Pykacz and Santos 1991; Beltrametti and M\c{a}czy\'{n}ski 1992; Pulmannov\'{a} and Majernik 1992; Dvure\v{c}enskij and L\"{a}nger 1995; Pulmannov\'{a} 1994, 2002) always contain terms interpreted as probabilities of conjunctions of propositions. In view of above-mentioned doubts concerning unrestricted treating of meets in OMLs as representing conjunctions of propositions, also interpretation of these inequalities becomes doubtful.

In order to get mathematical tools suitable for constructing virtual `joint' probabilities of pairs of non-compatible propositions in OMLs, N\'{a}n\'{a}siov\'{a} (2003) introduced a notion of an $s$-map (map for simultaneous measurements) subsequently studied in numerous papers (N\'{a}n\'{a}siov\'{a} and Khrennikov 2006, 2007; Al-Adilee and  N\'{a}n\'{a}siov\'{a} 2009;  N\'{a}n\'{a}siov\'{a} and Pulmannov\'{a} 2009;  N\'{a}n\'{a}siov\'{a} and Val\'{a}\v{s}kov\'{a} 2010a, 2010b). The aim of the present paper is to study Bell-type inequalities on OMLs in which probability of conjunction of propositions is modeled as a value that an $s$-map takes on a pair of these  propositions instead of a probability of their meet.

\section{BIVARIATE MAPS ON ORTHOMODULAR LATTICES}

 We recall that an \emph{orthomodular lattice} (OML) is a lattice $L$
 with $0_L$ and $1_L$ as the smallest and the greatest element,
 respectively, endowed with a unary operation $a\mapsto a'$ such that
 the following conditions are satisfied:
 \begin{enumerate}
 \item[(i)] $a'':=(a')'=a$;
 \item[(ii)] $a\leq b$ implies $b'\leq a'$;
 \item[(iii)] $a\vee a'=1_L$;
 \item[(iv)] $a\leq b$ implies $b=a\vee(a'\wedge b)$.
 \end{enumerate}

 Condition (iv) is called the \emph{orthomodular law}. Elements of an OML are traditionally called \emph{propositions}, although in the case when an OML is used as a basic structure of a generalized probability calculus, the name \emph{events} is more appropriate. If an OML $L$
 is closed under countable lattice operations, then $L$ is called a
 $\sigma$-\emph{orthomodular lattice} ($\sigma$-OML). However, since in Bell-type inequalities only finite meets are concerned, in this paper we shall regard only the most basic case. In the quantum
 logic approach to quantum theory, $\sigma$-OMLs are considered as
  mathematical models of quantum
 mechanical propositions (see, e.g., Beltrametti and Cassinelli 1981; Pt\'{a}k and Pulmannov\'{a} 1991). In the traditional
 Hilbert space approach, this OML is a lattice of
 projections $\cal L(\cal H)$ onto closed linear subspaces of the corresponding Hilbert space $\cal{H}$. More generally, the
 set of projections in every von Neumann algebra forms a complete OML (Hamhalter 2003).

 Let $L$ be an OML. Two elements $a,b\in L$ are called
 \emph{orthogonal} (denoted $a\perp b$) iff $a\leq b'$, and $a,b$ are
 called \emph{compatible} (denoted $a\leftrightarrow b$) iff
 $a=(a\wedge b)\vee(a\wedge b')$. Notice that two projections on a
 Hilbert space are orthogonal iff the product of them is zero, and
 compatible iff they commute.

 A \emph{probability measure} or \emph{state} on $L$ is a mapping $m:L\to [0,1]$ such that
 \begin{enumerate}
 \item[(i)] $m(1_L)=1$;
 \item[(ii)] $a\perp b$ implies $m(a\vee b)=m(a)+m(b)$.
 \end{enumerate}
 A state $m$ is \emph{$\sigma$-additive} if $m(a)=\sum_{i=1}^\infty m(a_i)$
 whenever $a=\bigvee_{i=1}^\infty a_i$ for any sequence $\{ a_i\}$ of
 pairwise orthogonal elements.

  Let $L$ be an OML. A map $p:L\times L \to [0,1]$ is called a \emph{map for simultaneous measurements} (abbr. \emph{s-map}) (N\'{a}n\'{a}siov\'{a} 2003) if the
 following conditions hold:
 \begin{enumerate}
 \item[(s1)] $p(1_L,1_L)=1$;
 \item[(s2)] if $a\perp b$, then $p(a,b)=0$;
 \item[(s3)]if $a\perp b$, then for any $c\in L$:
$$p(a\vee b,c)=p(a,c)+p(b,c),$$
$$p(c,a\vee b)=p(c,a)+p(c,b).$$

 \end{enumerate}

The following properties of $s$-maps proved in (N\'{a}n\'{a}siov\'{a} 2003) will be of utmost importance in our considerations:
\begin{enumerate}
\item[(N1)] The map $m_p :L \rightarrow [0,1]$ such that $m_p (a) = p(a,a) = p(1_L,a) = p(a,1_L)$ is a state on $L$. Such a state will be called a \emph{state generated by $p$}.
\item[(N2)] If $a \leftrightarrow b$, then $p(a,b) = m_p (a\wedge b) = p(b,a)$. This property shows that $s$-maps can be seen as providing probabilities of `virtual' conjunctions of propositions, even non-compatible ones, for in the case of compatible propositions ($a\leftrightarrow b$) the value $p(a,b)$ coincides with the value that a state $m_p$ generated by $p$ takes on the meet $a\wedge b$, which in this case really represents conjunction of $a$ and $b$.
\item[(N3)] $p(a',b') = 1-p(a,a) - p(b,b) + p(a,b)$  for all elements of an OML.
\end{enumerate}

Let us note that $s$-maps on OMLs resemble copulas (see, e.g., Nielsen 1999) that are used in classical probability and statistics to construct joint probability distributions from the given marginal probability distributions. 

It was shown many years ago by Greechie (1971) that there exist OMLs admitting no states. Of course also no $s$-map can be defined on such OMLs, otherwise the above-mentioned property of $s$-maps (N1) would not hold. On the other hand, there exist OMLs with abundance of $s$-maps, e.g., it was proved by N\'{a}n\'{a}siov\'{a} and Pulmannov\'{a} (2009) that any tracial state on a von Neumann algebra with no type $I_2$ direct summand generates an $s$-map on the lattice of projections of this algebra. Also from Propositions 1.1 and 2.1 of (N\'{a}n\'{a}siov\'{a} 2003) it follows that there exist a lot of $s$-maps on OMLs with unital sets of states. 

In general the problem of existence of $s$-maps on various OMLs is far from being settled and deserves further investigations. However, since authors of numerous papers in which Bell-type inequalities on OMLs are formulated with the use of meets never bother about the existence of probability measures they use, we shall adopt the same position w.r.t. $s$-maps.

In (N\'{a}n\'{a}siov\'{a} et. al. 2007; see also N\'{a}n\'{a}siov\'{a} and  Val\'{a}\v{s}kov\'{a} 2010) the following notion of a \emph{join map} (\emph{j-map}) on an OML has been introduced: 

 Let $L$ be an OML. A map $q:L\times L \to [0,1]$ is called a \emph{join map} (abbr. \emph{j-map}) if the
 following conditions hold:
 \begin{enumerate}
 \item[(j1)] $q(0_L,0_L)=0$,\quad  $q(1_L,1_L)=1$;
 \item[(j2)] if $a\perp b$, then $q(a,b) = q(a,a) +q(b,b);$
 \item[(j3)]if $a\perp b$, then for any $c\in L$:\\
 $$q(a\vee b,c)=q(a,c)+q(b,c)-q(c,c)$$
$$q(c,a\vee b)=q(c,a)+q(c,b)-q(c,c).$$
 \end{enumerate}

\noindent It was proved in (N\'{a}n\'{a}siov\'{a} and  Val\'{a}\v{s}kov\'{a} 2010)  that if $p$ is an $s$-map on an OML, then $q_p (a,b) = p(a,a) + p(b,b) - p(a,b) = m_p(a) + m_p(b) - p(a,b)$ is a $j$-map\footnote{It is easy to see that if $a\leftrightarrow b$, then $q_p(a,b)=m_p(a)+m_p(b)-m_p(a\wedge b)=m_p(a\vee b)$ which explains its name.}, i.e., it maps $L\times L$ into $[0,1]$. This fact will be used as a starighforward justification of the most basic of  Bell-type inequalities concerning $s$-maps on OMLs, that will be studied  in the next section.

The third map $d: L\times L \to [0,1]$ which will be useful in our considerations, in Boolean case is a probability of symmetric difference $a\bigtriangleup b = (a\wedge b')\vee (a'\wedge b)$ of two propositions. This is the reason for which it was called in (N\'{a}n\'{a}siov\'{a} and  Val\'{a}\v{s}kov\'{a} 2010) a \emph{difference map} or simply \emph{d-map}. It is defined by the following conditions:
\begin{itemize}
\item[(d1)] $d(a,a)=0$  for any $a\in L$, and $d(1_L,0_L)=d(0_L,1_L)=1$;
\item[(d2)]  if  $a\perp b$, then $d(a,b)=d(a,0_L)+d(0_L,b)$;
\item[(d3)]  if $a\perp b$, then for any $c \in L$:  \\
$$d(a\vee b,c)=d(a,c)+d(b,c)-d(0_L,c)$$
$$d(c,a\vee b)=d(c,a)+d(c,b)-d(c,0_L).$$
\end{itemize}

It was shown in (N\'{a}n\'{a}siov\'{a} and  Val\'{a}\v{s}kov\'{a} 2010) that as in the case of $j$-maps, each $s$-map $p$  on an OML induces a $d$-map $d_p$ by the formula: $d_p(a,b) = p(a,b')+p(a',b)$ (the reverse assertion is not true).

The following properties of the $d$-map $d_p$ induced by an  $s$-map $p$ will be used in the sequel:

\begin{lem} Let $p$ be an s-map on an OML and $d_p$ be a d-map induced by $p$. Then
 \item[(a)] $d_p(a,b)=p(a,a)+p(b,b)-2p(a,b)$;
\item[(b)] $d_p(a,0_L) = d_p(0_L,a) =p(a,a) = m_p(a).$\\
\end{lem}
\noindent \textbf{Proof.} 

To show (a) it is enough to notice that from the condition (s3) of the definition of an $s$-map and its property (N1) it follows that $p(a,a)=p(a,1_L)=p(a,b\vee b') = p(a,b)+p(a,b')$, so $p(a,b')=p(a,a) - p(a,b)$. Analogoulsly, $p(a',b) = p(b,b) - p(a,b)$. By inserting these differences into the definition of $d_p$ we get (a).

Equalities (b) follow from (a) and from the fact that for any $a$ in an OML, $a\perp 0_L$, which by the condition (s2) of the definition of an $s$-map implies that $p(0_L,0_L) = p(a,0_L) = 0$.\\
\noindent
$\Box$

\section{BELL-TYPE INEQUALITIES IN WHICH PROBABILITY OF CONJUNCTIONS IS MODELED BY AN \emph{s}-MAP}

We shall study $s$-map counterparts of the following Bell-type inequalities involving meets, that were studied already by Pitovsky (1989) (see also Pulmannov\'{a} 2002; Dvure\v{c}enskij and L\"{a}nger 1995):
\begin{enumerate}
\item[(B1)] $m(a) + m(b) -m(a \wedge b) \leq 1$
\item[(B2)] $m(a) + m(b) + m(c) - m(a \wedge b) - m(a\wedge c) - m(b\wedge c) \leq 1$
\item[(C1)] $m(b) + m(c) \geq m(a\wedge b) + m(b\wedge c) + m(c\wedge d) - m(a\wedge d)$
\item[(C2)] $m(a\wedge b) + m(b\wedge c) + m(c\wedge d) - m(a\wedge d) - m(b) - m(c) \geq -1.$
\end{enumerate}
Inequalities (B1) and (B2) are usually called inequalities of \emph{Bell-Wigner type} while (C1) and (C2) are usually called inequalities of  \emph{Clauser-Horne type}. All these inequalities are satisfied by any probability measure on a BA.

Our aim is to study analogs of these inequalities in which, because of poperties of $s$-maps mentioned in the previous section, probabilities of single events $m(a), m(b)$, etc. are replaced, respectively, by $p(a,a), p(b,b)$, etc., and probabilities of joint occurences of events are modeled by values that an $s$-map $p$ takes on pairs of events. This means that we shall study the following inequalities:
\begin{enumerate}
\item[(B1$'$)] $p(a,a) + p(b,b) -p(a,b) \leq 1$
\item[(B2$'$)] $p(a,a) + p(b,b) + p(c,c) - p(a,b) - p(a,c) - p(b,c) \leq 1$
\item[(C1$'$)] $p(b,b) + p(c,c) \geq p(a,b) + p(b,c) + p(c,d) - p(a,d)$
\item[(C2$'$)] $p(a,b) + p(b,c) + p(c,d) - p(a,d) - p(b,b) - p(c,c) \geq -1.$
\end{enumerate}

Let us note, that since $s$-maps are in general non-commutative, each term $m(x\wedge y)$  in inequalities (B1) -- (C2) yields two generalizations to $s$-maps: $p(x,y)$ and $p(y,x)$. Therefore, there are two $s$-map counterparts of inequality (B1), eight of (B2), and sixteen of (C1) and (C2), and inequalities (B1$'$) -- (C2$'$) form only a sample of them. However, we expect that this sample is a faithful representative of them all and we restrict our considerations to (B1$'$) -- (C2$'$). Moreover, it occurs that some particularly interesting problems concerning Bell-type inequalities for $s$-maps, that will be discussed in details in Section 4 appear only in the case of commuting $s$-maps, in which case inequalities (B1$'$) --  (C2$'$) are the unique $s$-map generalizations of inequalities (B1) -- (C2).

It is straightforward to see that the expression on the left-hand side of inequality (B1$'$) is a $j$-map generated by an $s$-map $p$. Since any $j$-map takes values in the interval $[0,1]$, we obtain 

\begin{prop}\label{p1}
 Inequality (B1\,$'$) is satisfied by all s-maps defined on an OML.
\end{prop}

This result is of utmost importance since it shows that $s$-maps, invented to describe probabilities of virtual `joint' occurence of non-compatible events, are more close to classical probabilities than values $m(a\wedge b)$ usually interpreted as probabilities of coincidence of events $a$ and $b$. We stress once more that if events $a$ and $b$ are incompatible, the value $m(a\wedge b)$ cannot be checked in any experiment, so it can be replaced by any value that a bivariate map $p: L\times L \rightarrow [0,1]$ takes on a pair $a,b\in L$, provided that $p(a,b) = m(a\wedge b)$ if $a\leftrightarrow b$. This requirement is met by $s$-maps. Actually, $s$-maps were invented just to meet this requirement! In view of this result, the fact that inequality (B1) may be violated by probability measures defined on an OML is for quantum physics of no importance at all, since it shows that this violation can occur only for non-compatible propositions, i.e., it can be never checked experimentally.

\subsection{Bell-Wigner inequality (B2$'$) for $s$-map is a triangle inequality for induced $d$-map}

In the case of compatible propositions ($a \leftrightarrow b$) the expression in the part  (a) of Lemma 1 takes the form: $d_p(a,b) = m_p(a) + m_p(b) - 2m_p(a\wedge b)$, therefore, if $a\leftrightarrow b$, $d_p(a,b)$ coincides with the notion of \emph{separation} of $a$ and $b$  defined in this way by Santos already in 1986. Santos in his paper (1986) proved that if an OML is a Boolean algebra, then separation fulfills triangle inequality, in fact it is a pseudometric on an OML\footnote{A mapping $d: L\times L \rightarrow [0,1]$ is a \emph{pseudometric} on $L$ if for all $a,b,c  \in L$:   $d(a,a)=0; d(a,b) = d(b,a); d(a,b) \leq d(a,c) + d(c,b).$}. Moreover, he proved that triangle inequality for a separation is equivalent to the Bell-type inequality (C1). Since our $d$-map  $d_p(a,b)$ coincides with Santos' separation for a state $m_p$  on all pairs of compatible propositions, it is interesting to study connections of Bell-type inequalities for an $s$-map $p(a,b)$  with triangle inequalities for the induced $d$-map $d_p(a,b)$.
In this respect we get the following:

\begin{lem}Let $L$ be an OML, let $p$ be an s-map on $L$, and $d_p$ be a d-map induced by $p$. Bell-Wigner inequality (B2\,$'$) for the s-map $p$ is equivalent to the following triangle inequality for the d-map  $d_p$: 
\begin{enumerate}
\item[\emph{(}$\bigtriangleup $\emph{)}] $d_p(a,c) \leq d_p(a,b') + d_p(b',c)$.
\end{enumerate}

Moreover, the Clauser-Horne type inequalities (C1\,$'$) and (C2\,$'$) are equivalent, respectively, to the following inequalities:
\begin{enumerate}
\item[\emph{(}$\bigtriangleup '$\emph{)}] $d_p(a,d) \leq d_p(a,b) + d_p(b,c) + d_p(c,d)$;
\item[\emph{(}$\bigtriangleup'' $\emph{)}] $d_p(a,b) + d_p(b,c) + d_p(c,d) \leq 2 + d_p(a,d)$.
\end{enumerate}
\end{lem}
\noindent
\textbf{Proof.} 

The proof of equivalence (B2$'$) $\Leftrightarrow $ ($\bigtriangleup $) goes through the following sequence of equivalent  inequalities, where the differences of the type $p(a,b') = p(a,a) - p(a,b)$, already used in the proof of Lemma 1, and also the fact that $p(b',b') = m_p(b') = 1 - m_p(b) = 1 - p(b,b)$ are applied:
$$  d_p(a,c) \leq   d_p(a,b')+d_p(b',c)$$
$$ p(a,a)+p(c,c) -2p(a,c) \leq (p(a,a) + p(b',b')-2p(a,b'))+(p(b',b')+
p(c,c)-2p(b',c)).$$
It means that
$$-2p(a,c)\leq 2p(b',b')-2p(a,b')- 2p(b',c)$$
$$p(a,b')+p(b',c)-p(a,c)-p(b',b')\leq 0$$
$$ (p(a,a)-p(a,b))+(p(c,c)-p(b,c)) -p(a,c)-(1-p(b,b))\leq 0$$
$$ p(a,a)+p(b,b)+p(c,c)-p(a,b)-p(b,c)-p(a,c)\leq 1.$$

The proof of equivalence (C1$'$) $\Leftrightarrow $ ($\bigtriangleup '$) also goes through the sequence of equivalent inequalities. Let us begin with ($\bigtriangleup '$):
$$ d_p(a,d)\leq  d_p(a,b)+d_p(b,c)+d_p(c,d).$$
By condition (a) of Lemma 1 this inequality is equivalent to the following inequality: 
$$
 -2p(a,d) \leq    2p(b,b)-2p(a,b)+2p(c,c) -2p(b,c)-2p(c,d).
  $$
   It means that
 $$2p(a,b)+2p(b,c)+2p(c,d)-2p(a,d) \leq 2p(b,b)+2p(c,c), $$ so
$$p(a,b)+p(b,c)+p(c,d)-p(a,d) \leq p(b,b)+p(c,c) .$$

The proof of equivalence (C2$'$) $\Leftrightarrow $ ($\bigtriangleup''$) is analogous.\\
\noindent
$\Box$
\begin{rem}
Let us note that equivalences  (C1\,$'$) $\Leftrightarrow $ ($\bigtriangleup '$) and (C2\,$'$) $\Leftrightarrow $ ($\bigtriangleup''$) are `faithful' ones, in the sense that the same elements appear on both sides of these equivalences. This does not happen in the case of equivalence (B2\,$'$) $\Leftrightarrow $ ($\bigtriangleup $), but of course the following corollary holds:
\end{rem}

\begin{coro} If one of the inequalities (B2\,$'$) or ($\bigtriangleup $) is satisfied by all triples of elements of an OML, the same happens for the other one.
\end{coro}

\subsection{Equivalence of Bell-type inequalities involving more than two elements}

According to Proposition 1 the simplest Bell-type inequality (B1$'$) that involves only two elements is always satisfied by any $s$-map defined on an OML. The following example shows that this does not have to happen in the case of Bell-type inequalities that involve more than two elements.

\begin{ex1} Let $L$ be a horizontal sum of three Boolean algebras $2^2$: 

\vspace{1cm}
\unitlength=0.8mm
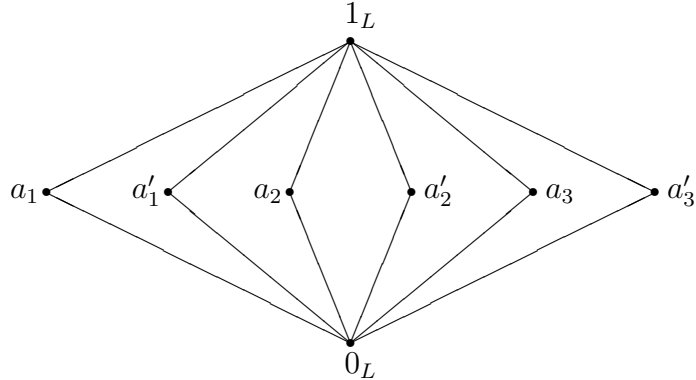
\begin{picture}(300,50)(0,0)
\put(75,0){\circle*{1.5}}
\put(75,50){\circle*{1.5}}
\put(25,25){\circle*{1.5}}
\put(45,25){\circle*{1.5}}
\put(65,25){\circle*{1.5}}
\put(85,25){\circle*{1.5}}
\put(105,25){\circle*{1.5}}
\put(125,25){\circle*{1.5}}
\put(75,0){\line(-2,1){50}}
\put(75,0){\line(-6,5){30}}
\put(75,0){\line(-2,5){10}}
\put(75,0){\line(2,5){10}}
\put(75,0){\line(6,5){30}}
\put(75,0){\line(2,1){50}}
\put(75,50){\line(-2,-1){50}}
\put(75,50){\line(-6,-5){30}}
\put(75,50){\line(-2,-5){10}}
\put(75,50){\line(2,-5){10}}
\put(75,50){\line(6,-5){30}}
\put(75,50){\line(2,-1){50}}
\put(19,24){\normalsize{$a_1$}}
\put(39,24){\normalsize{$a_1'$}}
\put(59,24){\normalsize{$a_2$}}
\put(87,24){\normalsize{$a_2'$}}
\put(107,24){\normalsize{$a_3$}}
\put(127,24){\normalsize{$a_3'$}}
\put(74,-5){\normalsize{$0_L$}}
\put(74,53){\normalsize{$1_L$}}
\end{picture}
\vspace{5mm}

\noindent  Let $p$ be commutative  bivariate map defined  on $L$  in the following way $ \forall i,j=1,2,3; \, i \neq j$:
\begin{eqnarray*}
p(1_L,1_L)&=&1,\quad  p(0_L,0_L)\; = \;0,\quad p(1_L,0_L)\; = \;0;\\
p(0_L,a_i)&=&p(0_L,a_i')\;=\;p(a_i,a_i')\;=\;0;\\ 
p(a_i,a_i)&=&p(a_i',a_i')\; =\; p(a_i,1_L)\; =\; p(a_i',1_L)\; =\; 0,5;\\ 
p(a_i,a_j)&=&0,1 ;\\
p(a_i,a_j')&=&p(a_i,a_i)-p(a_i,a_j\;)=\;0,4;\\
p(a_i',a_j')&=&1-p(a_i,a_i)-p(a_j,a_j)+p(a_i,a_j)\;=\;0,1.
\end{eqnarray*}
It is easy to check that $p$ is an s-map which, however, does  not satisfy Bell-Wigner inequality (B2\,$'$):
$$
p(a_1,a_1) + p(a_2,a_2) + p(a_3,a_3) - p(a_1,a_2) - p(a_2,a_3) - p(a_1,a_3) \leq 1$$
because
$$3\cdot 0,5 - 3\cdot 0,1 > 1. $$
\end{ex1}
However, due to results obtained in the previous subsection, we can prove the following:

\begin{prop} Bell-Wigner inequalities (B2\,$'$) and triangle inequalities ($\bigtriangleup $) are satisfied for all triples of elements of an OML if and only if inequalities  ($\bigtriangleup '$) and (C1\,$'$) are satisfied for all quadruples of elements of an OML. Moreover, if an s-map involved in these inequalities is commutative, also inequalities ($\bigtriangleup ''$) and (C2\,$'$) are satisfied.
\end{prop}
\noindent
\textbf{Proof.}  

According to Corollary 1 in this case we can write triangle inequality  ($\bigtriangleup $) in the usual form: $d_p(a,c) \leq d_p(a,b) + d_p(b,c)$. Then its equivalence with quadrilateral inequality ($\bigtriangleup '$) can be shown in a standard way: ($\bigtriangleup $) $\Rightarrow $ ($\bigtriangleup '$) because
$$ d_p(a,d) \leq d_p(a,b) + d_p(b,d) \leq d_p(a,b) + d_p(b,c) + d_p(c,d),$$
and the opposite implication is obtained by substitution $d \mapsto c$ in the quadrilateral inequality ($\bigtriangleup '$).

Since (C1$'$) $\Leftrightarrow $ ($\bigtriangleup '$) and (C2$'$) $\Leftrightarrow $ ($\bigtriangleup''$), to finish the proof it suffices to show that (C1$'$) $\Leftrightarrow$ (C2$'$) or ($\bigtriangleup '$) $\Leftrightarrow $ ($\bigtriangleup ''$). We shall show the first equivalence substituting $a \mapsto c$, $b\mapsto b'$, and $c\mapsto a$ in (C1$'$). Then, using the same substitutions as in previous proofs, we obtain a sequence of equivalent inequalities, where, however, commutativity of the $s$-map that appears in these inequalities, is utilized:
$$p(b',b')+p(a,a) \geq p(c,b')+p(b',a)+p(a,d)-p(c,d) $$
$$1-p(b,b)+p(a,a)-(p(c,c)-p(c,b))-(p(a,a)-p(a,b))-p(a,d)+p(c,d)\geq 0$$
$$-p(b,b)+p(a,a)-p(c,c)+p(c,b)-p(a,a)+p(a,b)-p(a,d)+p(c,d)\geq-1 $$
$$-p(b,b)-p(c,c)+p(a,b)+p(b,c)+p(c,d)-p(a,d)\geq -1$$
\noindent
$\Box $

\section{BELL-TYPE INEQUALITIES AND EXISTENCE OF MULTIVARIATE \emph{s}-MAPS}

In (N\'{a}n\'{a}siov\'{a} and Khrennikov 2007) the notion of an $s$-map was generalized to the notion of \emph{n-variate s-map} (abbr. \emph{s$_n$-map}) in the following way: 

Let $L$ be an OML. A map $p_n : L^n \rightarrow [0,1]$ is called an $s_n$-map if the following conditions hold:
\begin{itemize}
\item[(s$_n$1)] $p_n(1_L,...,1_L)=1$;
\item[(s$_n$2)]  if  $a_i\perp a_j$ for some $i\neq j$,  then $p_n(a_1,...,a_n)=0$;
\item[(s$_n$3)]  if $a_i\perp b_i$ for some $i$,  then  for all $c_1,...,c_{i-1}, c_{i+1},...,c_n \in L$
$$
p_n(c_1,...,a_i\vee b_i,...,c_n)=
p_n(c_1,...,a_i,...,c_n)+p_n(c_1,..., b_i,...,c_n). 
$$
\end{itemize}

The aim of introducing $s_n$-maps was to construct joint probability distributions of more than two non-compatible observables. The properties of multivariate $s_n$-maps are the same as properties of bivariate $s$-maps (which, actually, are $s_2$-maps).

Loosely speaking, in classical probability theory from any joint probability distribution one can obtain marginal probability distributions by replacing some random events by the sure event $\Omega $. Inspired by this,  N\'{a}n\'{a}siov\'{a} and Khrennikov (2007) defined, for a given $s_n$-map $p_n$ and every $k <  n$, a \emph{marginal $s_k$-map} by the formula:
$$p_k: L^k\rightarrow [0, 1],\quad  p_k(a_1,a_2,...,a_k)=p(\underbrace{a_1,a_2,...,a_k,1_L,...,1_L}_{n}).$$ 

Although in general  $s_n$-maps are not invariant with respect to permutations, they proved  that this happens  when any two arguments of an $s_n$-map are compatible. Therefore, in the definition of a marginal $s_k$-map the maximal element $1_L$ of an OML can be placed at any position. They also proved that an $s_k$-map that   is a marginal map of some $s_n$-map with $n > k$ is always invariant with respect to permutations. In particular, this means that in  the case of bivariate $s$-maps, all of them that are marginal $s_2$-maps of some $s_3$-maps, are commutative.

In numerous papers written by members of the so called `probabilistic opposition'\footnote{A term coined by Khrennikov and used by him, e.g.,  in (2009).} to the usual interpretation of violation of Bell-type inequalities (see (Khrennikov 2009) and references cited therein),  violation of Bell-type inequalities is not ascribed to nonlocality or lack of realism, but rather results from non-existence of a joint probability distribution that could yield marginal distributions being in accordance with probabilities obtained from quantum-mechanical calculations. In our `meet-free' approach we get the following result: 
\begin{prop}
Let $L$ be an OML and  let $p$ be a bivariate commutative $s_2$-map on $L$. If  for some $a,b,c \in L$ Bell-Wigner inequality (B2\,$'$)
is not satisfied, then trivariate $s_3$-map with marginal bivariate $s_2$-map $p$ does not exist. On the other hand, if a trivariate $s_3$-map $p'$  exists, then Bell-Wigner inequality (B2\,$'$) is satisfied by all marginal $s_2$-maps obtained from $p'$.
\end{prop}
\noindent
\textbf{Proof.}
If Bell-Wigner inequality (B2\,$'$)
is not satisfied for some $a,b,c \in L$, then
$1-p(a,a) - p(b,b) - p(c,c) + p(a,b) + p(a,c) + p(b,c) < 0$.
 Let us assume that  trivariate $s$-map $p_3$ with marginal bivariate $s$-map $p$ exists. Then  
$p_3(x,y,1_L)=p_3(x,1_L,y)=p_3(1_L,x,y)=p(x,y)$. 
Let us denote $p_3(a,b,c) = \alpha \in [0,1]$. Using definition of marginal $s_2$-maps and property (N3) of $s$-maps we get:
\begin{eqnarray*}
 p_3(a',b',c')&=&p_3(a',b',1_L)-p_3(a',b',c)\\
&=&p(a',b')-(p_3(a',1_L,c)-p_3(a',b,c))\\
&=&p(a',b')-p(a',c)+(p_3(1_L,b,c)-p_3(a,b,c))\\
&=&p(a',b')-p(a',c)+p(b,c)-\alpha\\
&=&(1-p(a,a) - p(b,b) + p(a,b)) -(p(c,c)-p(a,c))+p(b,c)-\alpha\\
&=&1-p(a,a) - p(b,b) - p(c,c) + p(a,b) + p(a,c) + p(b,c)-\alpha.
\end{eqnarray*}
Therefore, the value of $p_3(a',b',c')$ is negative for any $\alpha \in [0,1]$, which is impossible.

The second part of Proposition follows directly from its first part.

\noindent
$\Box $

Proposition 3 shows that even within the, we dare say, more correct approach, in which probabilities of conjunctions of propositions are not calculated as values that probability measures defined on OMLs take on meets of elements of an OML, but rather as values that $s$-maps take on pairs of these  propositions, violation of Bell-Wigner inequality (B2\,$'$) means that a (generalized) joint probability distribution that could be used to describe the experimental situation does not exist.

\section{THE LOGIC OF COUNTERFACTUAL PROPOSITIONS ABOUT QUANTUM SYSTEMS}

As it was mentioned in the Introduction, in Boolean algebras which are algebraic representation of families of experimentally verifiable propositions pertaining to classical physical systems, meets and joins are proper models of conjunctions and disjunctions of propositions. However, unrestricted generalization of this statement to OMLs that are algebraic representations of `quantum logics', i.e., sets of experimentally verifiable propositions pertaining to quantum systems, leads to numerous difficulties caused by the fact that it is not possible to verify simultaneously propositions which are represented by non-compatible elements of an OML. (In order to simplify the language such propositions about quantum systems will be themselves called \emph{non-compatible} in the sequel. We shall also often identify propositions about quantum system with elements of an OML that represent them). Actually, according to the  strict `verificationist' point of view, conjunctions and disjunctions of non-compatible propositions should be regarded as meaningless.

According to the traditional approach, originated by Birkhoff and von Neumann in their historic paper (1936) `quantum logic' is regarded as 2-valued logic, which is non-classical because of non-distributivity. However, one of the authors in a series of papers (see, e.g., Pykacz 1994, 2000, 2010) promoted an idea that `quantum logic' can be equivalently regarded as a specific $\infty$-valued {\L}ukasiewicz logic, which opens the possibility of working out a new interpretation of quantum mechanics (Pykacz 2011). In this approach conjunctions and disjunctions of propositions about quantum systems are modelled by a pair of partially defined operations used in a  specific version of {\L}ukasiewicz many-valued logic. However, when they are defined, they necessarily coincide with meets and joins (Pykacz, 2000). Therefore, similarly to lattice operations of meet and join, they cannot be treated as proper models of conjunctions and disjunctions of non-compatible propositions.

The notion of an $s$-map opens a new possibility: if propositions $a$ and $b$  are non-compatible, the value $p(a,b)$ can be thought of as representing probability of simultaneous verification of $a$ and $b$ in a `counterfactual measurement': \emph{`what would be the probability of simultaneous verification of propositions  $a$ and $b$ if we were able to perform it'} or, according to the approach propounded in (Pykacz 1994, 2000, 2010) \emph{`what would be the truth-value of the ``counterfactual conjunction" of propositions a and b'}. Let us remind that the traditional belief that $m(a \wedge b)$ always represents probability of simultaneous  verification of propositions $a$ and $b$ is based on the tacit, and in the case of non-compatible propositions  erroneous assumption, that  this simultaneous verification is always possible.

Similarly, the value $q_p (a,b) = p(a,a) + p(b,a,) - p(a,b)$ can be thought of as representing truth-value of `counterfactual disjunction' of propositions represented by $a$ and $b$.

Therefore, we see that an $s$-map $p: L\times L \rightarrow [0,1]$ and its associated $j$-map $q_p$ allow to define for a studied quantum system a kind of a `logic of counterfactuals' in which truth-values of a \emph{counterfactual conjunction} $a {\bigtriangleup}_p b$ and a \emph{counterfactual disjunction}  $a {\bigtriangledown}_p b$ of two non-compatible propositions are modelled by $p(a,b)$ and $q_p(a,b)$, respectively.

Let us note that
$$
p(a',b')= 1-p(a,a)-pp(a,b)+p(a,b) = 1-q_p (a,b)
$$
and
\begin{eqnarray*}
q_p(a',b')&=& p(a',a') + p(b',b') - p(a',b')\\
&=& 1-p(a,a)+1-p(b,b) - 1 + p(a,a) + p(b,b) - p(a,b)\\
&=& 1-p(a,b).
\end{eqnarray*}
If we assume, as usual, that orthocomplementation in an OML represents logical negation and, as it is always assumed in {\L}ukasiewicz many-valued logic (1970), that truth-values of a proposition and its negation sum up to 1, we recognize in the formulas written above numerical expressions of both De Morgan laws:
$$ a' {\bigtriangleup}_p b' = [a {\bigtriangledown}_p b]', $$
$$ a' {\bigtriangledown}_p b' = [a {\bigtriangleup}_p b]'. $$
Since conjunction and disjunction of compatible propositions are properly modelled by their meet and join, the validity of numerous other laws, like the law of excluded middle, the law of contradiction and the orthomodular law, is secured by properties of meet and join in OMLs. Therefore, using $s$-maps and associated with them $j$-maps, we have obtained a kind of an `extended quantum logic' in which conjunction and disjunction is meaningful both in the case of compatible, as well as non-compatible propositions.

\section{SUMMARY}

Since interpretation of meets as representing conjunctions of propositions about quantum-mechanical systems is doubtful when these propositions are non-compatible, we studied various Bell-type inequalities on OMLs in which probabilities of meets of propositions were replaced by values that an $s$-map -- an object invented to model probabilities of simultaneous measurements of incompatible propositions -- takes on these propositions. It is significant that although the simplest Bell-type inequality (B1): $m(a) + m(b) - m(a\wedge b) \leq1$ may be violated on an OML (of course only by non-compatible elements), its $s$-map counterpart (B1$'$): $p(a,a) + p(b,b) -p(a,b) \leq 1$ is always satisfied. This shows that  replacing $m(a\wedge b)$ by $p(a,b)$ brings us closer to the situation encountered in classical probability theory, hopefully also closer to reality.

Nevertheless, Proposition 3 shows that even within our approach violation of Bell-Wigner inequality (B2\,$'$) by an $s_2$-map $p$ means that there does not exist an $s_3$-map for which $p$ would be a marginal $s_2$-map. This is in accordance with numerous papers written by `probabilistic opposition' to the usual interpretation of violation of Bell-type inequalities, in which violation of Bell-type inequalities is not ascribed to non-existence of `local realism', but rather indicates impossibility of constructing a single probability space in which experiments designed to check Bell-type inequalities could be embedded. Whether this impossibility follows from, or is equivalent to, the non-existence of `local realism', should be the aim of further investigations.

Finally, we showed that one can treat values that an $s$-map $p$ takes on non-compatible propositions about quantum systems as truth-values of `counterfactual conjunctions' of these propositions, and similarly values that an associated $j$-map $q_p$ takes on such propositions as truth-values of `counterfactual disjunctions' of them. This allows to construct propositional calculus (`extended quantum logic') in which conjunctions and disjunctions of both compatible and non-compatible propositions are meaningful.

\medskip

\noindent
\textbf{Acknowledgments.} JP gratefully acknowledges financial support of the Polish National Center for Science (NCN) under the grant 2011/03/B/HS1/04573 and a scholarship of the National Scholarship Programme of the Slovak Republic which enabled him to stay three months in Bratislava and work on this paper. ON and LV acknowledge financial support by the grants VEGA 1/0143/11 and VEGA 1/0297/11.


\begin{thebibliography}{99}

\bibitem{AN07} Al-Adilee, A., N\'{a}n\'{a}siov\'{a}, O.: Copula and s-map on a quantum logic. Information Sciences. 179, 4199--4207 (2009)

\bibitem{BC81} Beltrametti, E., Cassinelli, G.: The Logic of Quantum Mechanics. Addison-Wesley, Reading MA (1981)

\bibitem{BM92} Beltrametti, E., M\c{a}czy\'{n}ski, M.: Problem of classical and nonclassical probabilities. Int. J. Theor. Phys. 31, 1849--1856 (1992)

\bibitem{BvN36} Birkhoff, G., von Neumann, J.: The logic of quantum mechanics. Annals of Mathematics. 37, 823--843 (1936)

\bibitem{DL95}  Dvure\v{c}enskij, A., L\"{a}nger, H.: Bell-type inequalities in orthomodular lattices, I-II. Int. J. Theor. Phys. 34, 995--1024, 1025--1036 (1995)

\bibitem{Gre71} Greechie, R.: Orthomodular lattices admitting no states. Journal of Combinatorial Theory. 10, 119--132 (1971)

\bibitem{Ham03} Hamhalter, J.: Quantum Measure Theory. Kluwer, Dordrecht (2003)

\bibitem{Jam74} Jammer, M.: The Philosophy of Quantum Mechanics. Wiley-Interscience, New York (1974)

\bibitem{Khr09} Khrennikov, A.: Violation of Bell's inequality and non-Kolmogorovness. In:  Accardi, L., et. al. (eds.) Foundations of Probability and Physics - 5.  American Institute of Physics, Mellville NY (2009)

\bibitem{Luk70} {\L}ukasiewicz, J.: Selected Works. North Holland, Amsterdam (1970)

\bibitem{Nan03} N\'{a}n\'{a}siov\'{a}, O.: Map for simultaneous measurements for a quantum logic. Int. J. Theor. Phys.  42, 1889--1903 (2003)

\bibitem{NK06} N\'{a}n\'{a}siov\'{a}, O., Khrennikov, A.: Representation theorem for observables on a quantum system. Int. J. Theor. Phys. 45, 481-494 (2006)

\bibitem{NK07} N\'{a}n\'{a}siov\'{a}, O., Khrennikov, A.: Compatibility and marginality. Int. J. Theor. Phys. 46, 1083--1095 (2007)

\bibitem{NP09} N\'{a}n\'{a}siov\'{a}, O.,  Pulmannov\'{a}, S.: S-map and tracial states. Information Sciences. 179, 515--520 (2009)

\bibitem{NV10a} N\'{a}n\'{a}siov\'{a}, O., Val\'{a}\v{s}kov\'{a}, L.: Maps on a quantum logic. Soft Computing. 14, 1047--1052  (2010)

\bibitem{NV10b} N\'{a}n\'{a}siov\'{a}, O., Val\'{a}\v{s}kov\'{a}, L.: Marginality and triangle inequality. Int. J. Theor. Phys. 49, 3199--3208 (2010)

\bibitem{NMM07} N\'{a}n\'{a}siov\'{a}, O.,  Min\'{a}rov\'{a}, M., Mohammed, A.: Probability and quantum logic. Forum Statisticum Slovacum. 5/2007, 101--107 (2007)

\bibitem{Nie99} Nielsen, R.B.: An Introduction to Copulas. Springer, New York (1999)

\bibitem{Pit89} Pitovsky, I.: Quantum Probability -- Quantum Logic. Springer, Berlin (1989)

\bibitem{PP91} Pt\'{a}k, P.,  Pulmannov\'{a}, S.: Orthomodular Structures as Quantum Logics. Kluwer, Dordrecht (1991)

\bibitem{Pul94} Pulmannov\'{a}, S.: Bell inequalities and quantum logics. In: Accardi, L. (ed.) The Interpretations of Quantum Theory: Where Do We Stand? Encyclopedia Italiana, Roma (1994)

\bibitem{Pul02} Pulmannov\'{a}, S.: Hidden variables and Bell inequalities on quantum logics. Foundations of Physics. 32, 193--216 (2002)

\bibitem{PM92} Pulmannov\'{a}, S., Majernik, V.: Bell inequalities on quantum logics. J. Math. Phys. 33, 2173--2178 (1992)

\bibitem{Pyk89} Pykacz, J.: On Bell type inequalities in quantum logics. In: Bitsakis, E.J., Nicolaides, C.A. (eds.) The Concept of Probability. Kluwer, Dordrecht (1989)

\bibitem{Pyk94} Pykacz, J.: Fuzzy quantum logic and infinite-valued {\L}ukasiewicz logic. Int. J. Theor. Phys. 33, 1403--1416 (1994)

\bibitem{Pyk00} Pykacz, J.: {\L}ukasiewicz operations in fuzzy set and many-valued representations of quantum logics. Foundations of Physics. 30, 1503 --1524 (2000)

\bibitem{Pyk10} Pykacz, J.: Unification of two approaches to quantum logic: Every Birkhoff -- von Neumann quantum logic is a partial infinite-valued {\L}ukasiewicz logic. Studia Logica. 95, 5--20 (2010)

\bibitem{Pyk11} Pykacz, J.: Towards many-valued/fuzzy interpretation of quantum mechanics. Int. J. Gen. Syst. 40, 11--21 (2011)

\bibitem{PS91} Pykacz, J., Santos, E.: Hidden variables in quantum logic approach revisited. J. Math. Phys. 32, 1287--1292 (1991)

\bibitem{San86} Santos, E.: The Bell inequalities as tests of classical logic. Physics Letters A. 115, 363--365  (1986)

\end{thebibliography}
\end{document}